\documentclass[newstyle,guide]{rmaa}
\usepackage{paralist}
\usepackage[latin1]{inputenc}

\title{CCD $U\!BV\!RI$ Photometry of the Galactic open clusters\altaffilmark{1}: Be~89, Ru~135, and Be~10}

\author{
{\.Inci} Akkaya\altaffilmark{2}, 
William J. Schuster\altaffilmark{3}, 
Ra\'ul Michel\altaffilmark{3},
Carlos Chavarr\'{\i}a--K\altaffilmark{3},
Andr\'e  Moitinho\altaffilmark{4}, 
Roberto V\'azquez\altaffilmark{3},
Y\"uksel Karata\c{s}\altaffilmark{5}
}

\altaffiltext{1}{based on observations carried on at the San Pedro M\'artir 
National Astronomical Observatory (SPM), operated by Instituto de 
Astronom\'{\i}a/Campus-Ensenada, Universidad Nacional Aut\'onoma de M\'exico.} 

\altaffiltext{2}{Department of Astronomy and Space Sciences, Erciyes University, Kayseri, Turkey}

\altaffiltext{3}{Instituto de Astronom\'{\i}a, Universidad Nacional Aut\'onoma de M\'exico, Campus-Ensenada, B.C., M\'exico}

\altaffiltext{4}{SIM/IDL, Facultade de Ciencias da Universidade de Lisboa, Lisboa, Portugal}

\altaffiltext{5}{Istanbul University, Science Faculty, Department of Astronomy and Space Sciences, Turkey}

\shortauthor{\.I. Akkaya et al.}

\shorttitle{$U\!BV\!RI$ Photometry of open clusters}

\fulladdresses{
\item \.Inci Akkaya:  Department of Astronomy and Space Sciences, Faculty of Arts and Sciences,
Erciyes University, Talas Yolu, 38039, Kayseri, Turkey (iakkaya@erciyes.edu.tr).

\item William J. Schuster, Ra\'ul Michel, Carlos Chavarr\'{\i}a--K, and Roberto V\'azquez:
Observatorio Astron\'omico Nacional, Instituto de Astronom\'{\i}a, Universidad Nacional
Aut\'onoma de M\'exico, Apartado Postal 877, C.P. 22800, Ensenada, B.C., M\'exico
(schuster, rmm, chavarri, vazquez@astrosen.unam.mx).

\item Andr\'e  Moitinho:  SIM/IDL, Facultade de Ciencias da Universidade de Lisboa, Ed. C8,
 Campo Grande, 1749-016, Lisboa, Portugal (andre@sim.ul.pt).

\item Y\"uksel Karata\c{s}:  Istanbul University, Science Faculty, Department of Astronomy and 
Space Sciences, 34119, \"Universite-Istanbul, Turkey (karatas@istanbul.edu.tr).

             }

\resumen{Presentamos los par\'ametros fundamentales de enrojecimiento,
metalicidad, edad y distancia de los c\'umulos abiertos poco estudiados
Be~89, Ru~135 y Be~10, derivados de la fotometr\'{\i}a CCD $U\!BV\!RI$
de los mismos.  Las distancias y edades de los c\'umulos
se obtuvieron ajustando las isocronas apropiadas a las secuencias observadas
de los c\'umulos para cinco diferentes diagramas color--magnitud.  Los promedios
pesados de m\'odulos de distancia y de distancias helioc\'entricas,
($(V_0$--$M_{V}), d$(kpc)) encontrados son los siguientes: $(11\fm90\pm
0\fm06, 2.4\pm 0.06$) para Be~89, $(9\fm58\pm 0\fm07, 0.81\pm 0.03$) para Ru~135 y
$(11\fm16\pm 0\fm06, 1.7 \pm 0.05$) para Be~10, mientras que los promedios pesados
para las edades $(\log(A), A$(Gyr)) son $(9.58\pm 0.06, 3.8\pm 0.6$)
para Be~89, $(9.58\pm 0.06, 3.8\pm 0.7$) para Ru~135 y $(9.06\pm 0.05, 1.08\pm 0.08$)
para Be~10.} 

\abstract{
The fundamental parameters of reddening, metallicity, age, and distance are presented
for the poorly studied open clusters Be~89, Ru~135, and Be~10, derived from their CCD
$U\!BV\!RI$ photometry.  By fitting the appropriate isochrones to the observed
sequences of the clusters in five different color--magnitude diagrams, the weighted
averages of distance moduli and heliocentric distances ($(V_0$--$M_{V}), d$(kpc)) are
$(11\fm90\pm 0\fm06, 2.4\pm 0.06$) for Be~89, $(9\fm58\pm 0\fm07, 0.81\pm 0.03$) for
Ru~135, and $(11\fm16\pm 0\fm06, 1.7 \pm 0.05$) for Be~10, and the weighted averages
of the ages $(\log(A), A$(Gyr)) are $(9.58\pm 0.06, 3.8\pm 0.6)$ for Be~89, $(9.58\pm
0.06, 3.8\pm 0.7)$ for Ru~135, and $(9.06\pm 0.05, 1.08\pm 0.08)$ for Be~10.

~ 

     } 
\addkeyword{CCD UBVRI photometry}
\addkeyword{open clusters and associations: individual (Be10, Be89, Ru135)}
\addkeyword{stars: Hertzsprung-Russell and C--M diagrams}
\addkeyword{stars: fundamental parameters}

\begin{document}

\maketitle

\section{Introduction}  
\label{sec:intro}

Galactic open clusters, which contain a few tens to a few tens of thousands
of stars and are a few parsecs across, are sparsely populated, loosely
concentrated, and gravitationally bound systems.  With systematic image searches
and follow-up photometric surveys, new open clusters are currently being
discovered.  By fitting the photometric observations of open clusters to synthetic
photometry resulting from stellar models (i.e.~theoretical isochrones), which
include the newest input physics, stellar structure, and differing heavy-element
abundances, fundamental parameters such as interstellar reddening, metallicity,
distance modulus, and age can be precisely and accurately {\bf determined}.  These
parameters have great importance concerning the age--metallicity relation and
the metal-abundance gradient in the Galactic disk \citep[e.g.][]{cameron85,
carraro94,edfriel95}, as well as the luminosity and mass functions of the
open clusters \citep{pisetal08}. Open clusters are also very useful for
testing the stellar evolutionary models, given that their stars were formed
at the same time, out of the same cloud, and under similar environmental
conditions. Thus, open clusters are ideal entities for the study of stellar
evolution since physical properties are tightly constrained, being mainly
distinguished by the stellar mass, so that theoretical models of stellar
formation and evolution can be compared with real clusters without excessive
complications.  For these analyses, the fundamental parameters such as
interstellar reddening, metallicity, distance modulus, and age should be
determined as precisely and accurately as possible.

In Galactic studies, one of the more severe observational limitations is
due to the absence of photometric data for nearly half of the approximately
1500 open clusters known. Furthermore, there is a lack of homogeneity in the
observations and analyses of the clusters studied. The catalogue of Lyng{\aa}
(1987), that resulted from a collection of data from many different sources and
which includes 422 open clusters, constituted the observational basis for a
large number of astronomical studies, led to important conclusions about
the Galactic disk, and has been very useful for planning subsequent
observations by other astronomers.  However, this catalogue has been built
from parameters obtained by various authors, with diverse observing techniques,
distinct calibrations, and different criteria for determining the stellar
ages, rendering it very inhomogeneous and limited for studies requiring
precision in the measurement of these fundamental parameters.  As an example
of the precision and accuracy that one can expect due to the effects of these
inhomogeneities, we refer to Janes \& Adler (1982), who found that distance
moduli of a given cluster obtained by two or more authors have a mean
difference of $0\fm55$.

Within the Sierra San Pedro M\'artir, National Astronomical Observatory (hereafter
SPM) open cluster project \citep[cf.][]{schetal07,micetal10}, the aims are
the following:
\begin{enumerate}
 \item A common {\sl UBVRI} photometric scale for open clusters,
 \item An atlas of color--color and color--magnitude diagrams for these clusters,
 \item A homogeneous set of cluster reddenings, distances, ages, and, if possible, metallicities,
 \item An increased number of old, significantly reddened, or distant, open clusters, and
 \item A selection of interesting clusters for further study.
\end{enumerate}
The open clusters for the present study were selected from the large (and most complete)
catalogue , ``Optically visible  open Clusters and Candidates'' \citep{diaetal02}, which
is now also available at the CDS  \footnote[6]{{\bf http://www.astro.iag.usp.br/$\sim$wilton/}}
(Centre de Donn\'ees Astronomiques de Strasbourg). This work aims to provide the fundamental parameters of reddening, metallicity, distance modulus
and age for the open clusters Be~89, Ru~135, and Be~10.  Our final intention is to publish a
set of homogeneous photometric $U\!BV\!RI$ data for over 300 Galactic clusters \citep{schetal07,
tapetal10}
 
This paper is organized as follows:  $\S2$ describes the observations and reduction techniques.
$\S3$ contains the derivation from the $U\!BV\!RI$ photometry of reddening and metallicity of
the clusters from two--color diagrams, and the inference of distance moduli and ages from
color--magnitude diagrams. Their uncertainties are also discussed. Comparisons of these
parameters with previous results from the literature are made in $\S4$, and the conclusions
are given in $\S5$.

\begin{figure}[!p] 
\includegraphics[width=2.35\columnwidth]{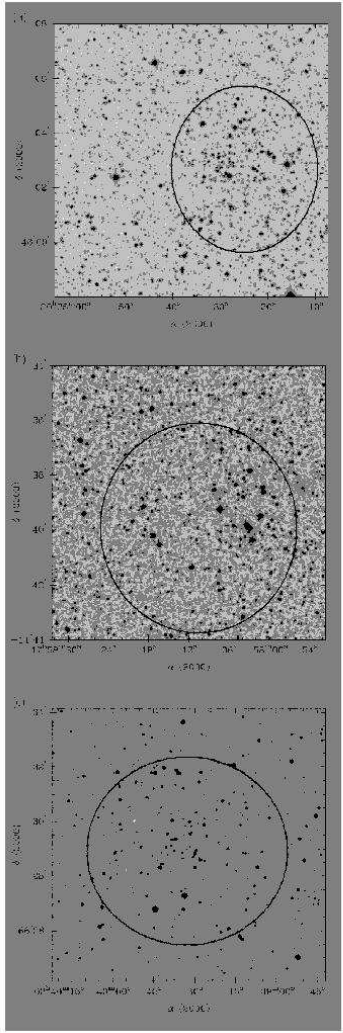}
\caption{
DSS red-filter images of the Galactic open clusters Be~89 (panel~a),
Ru~135 (b), and Be~10 (c). The regions analyzed with the {\sl ELIPSE} inspection tool, to derive
first estimates for the fundamental parameters, are enclosed by ellipses.  Orientation as usual:
north is up, and east to the left.
        }
\label{fig:DSS}
\end{figure} 

\section{Observation and reduction techniques} 

\subsection{The observations}  

This CCD {\sl UBVRI} project of northern open clusters has been undertaken at SPM using
always the same instrumental setup (telescope, CCD-detector, and filters), observing 
procedures, reduction methods, and system of standard stars (Landolt 1983, 1992).  A
par focal set of {\sl UBVRI} Johnson-Cousins filters was used for our observations.
The 0.84-m f/13 Cassegrain telescope hosted the filter-wheel ``Mexman'' provided with
the SITE\#1 (SI003) CCD camera, which has a 1024$\times$1024 square pixel array and a
$24\mu$m$\times24\mu$m pixel size; this CCD has non-linearities less than 0.45\% over a
wide dynamical range, no evidence for fringing even in the {\sl I} band, and Metachrome
II and VISAR coverings to increase sensitivity in the blue and near ultraviolet.  The
sky-projected pixel size was 0\farcs 393, and the field of view of the detector was
$6.73\times 6.73 \;$ arcmin$^2$.  Here the results of {\sl UBVRI} images for the open
clusters Be~89, Ru~135, and Be~10 are presented, which were acquired in July 2001 (Be~89
and Ru~135) and February 2002 (Be~10).  The exposure times were typically $3 \times 240$
seconds for the $U$ filter, $3 \times 180$ for $B$, $3 \times 100$ for $V$, $3 \times 100$
for $R$, and $3 \times 120$ for $I$.  Several standard-star fields from Landolt (1992)
were observed nightly to permit the derivation of the photometric transformations to the
Johnson-Cousins' system and the atmospheric extinction coefficients.  For the July 2001
observing run, seven Landolt groups were used, containing 26 different standard stars
with color ranges, $-0\fm25 \le (B$--$V) \le +1\fm14$, $-1\fm09 \le (U$--$B) \le +1\fm14$, and
$-0\fm30 \le (V$--$I) \le +1\fm14$.  Sixteen to twenty-five observations of these Landolt
standards were made per night.  For the February 2002 run, eight Landolt groups were
employed, containing 35 different standard stars with color ranges, $-0\fm30 \le (B$--$V)
\le +1\fm42, -1\fm18\le (U$--$B)\le +1\fm27$, and $-0\fm28 \le (V$--$I) \le +1\fm77$.
Fifty-two to seventy-two observations of these Landolt standards were made per night,
except one night cut short by clouds, when only 15 observations were managed.  The
standard-star fields have been observed with exposures of $1 \times 240$ seconds for the
{\sl U} filter, $1 \times 120$ for {\sl B}, $1 \times 60$ for {\sl V}, $1 \times 60$ for
{\sl R}, and $1 \times 60$ for {\sl I}.  The observed Landolt fields and the number of
associated stars in each one are summarized in Table~1.

\begin{table}[!htbp] 
\caption{L\lowercase {andolt's fields of standard stars}}
   {\scriptsize 
\begin{center}
\begin{tabular}{lclc} 
\hline  
\multicolumn{2}{c}{July 2001}&\multicolumn{2}{c}{February 2002} \\[2mm] 
\multicolumn{1}{c}{region} & N$_{stds}$& \multicolumn{1}{c}{region} & N$_{stds}$ \\
\hline 
 PG1528+062 & 3~~~ & ~~PG0918+029 & 5 \\ 
 PG1530+057 & 3~~~ & ~~PG0942-029 & 5 \\ 
 PG1633+099 & 5~~~ & ~~PG1047+003 & 4 \\ 
 PG1657+078 & 4~~~ & ~~PG1323-086 & 5 \\ 
 PG2213-006 & 4~~~ & ~~PG1528+062 & 3 \\ 
 PG2331+055 & 3~~~ & ~~SA~95      & 5 \\ 
 MARK~A     & 4~~~ & ~~SA104      & 4 \\ 
            &      & ~~SA107      & 4 \\[2mm] 
 ~~~~TOTAL  &26~~~ & ~~~~~~TOTAL  &35 \\ \hline
\end{tabular} 
\end{center} 
     }
\end{table}  

Usually one or more Landolt fields were re-observed nightly with an air-mass range of
at least 0.70 in order to measure the coefficients of the atmospheric extinction of the
SPM site, which has excellent sky conditions.  To improve the accuracy, precision, and 
efficiency of the photometric observations when required, the filters were observed in
forward and backward sequences (i.e.\ {\sl UBVRI-IRVBU}), especially for the large air-mass
observations.

\subsection{Data reduction}  

The usual (night and run) calibrations for CCD photometry were done during each of our
observing periods (i.e.\ bias, twilight-sky flat fields, and dark-current determinations) to
determine the (night and run) mean correcting frames.  Standard data reduction procedures
have been used within {\it IRAF},\footnote[5]{{\it IRAF} is distributed by NOAO (operated by the
Association of Universities for Research in Astronomy, Inc.) under cooperative agreement
with NSF} the CCDRED and DAOPHOT tasks
\citep[aperture and PSF photometry, see][]{showell89,showell90,stetson87,stetson90}.
More details concerning the instrumentation and the observing and reduction
procedures of this project will be given in the near future in the succeeding paper of this
project \citep[][and references therein]{micetal10}.  
To obtain the magnitudes and colors on the standard system for the stars 
associated with these clusters, we followed  Jordi et al.\ (1995), and
Rossell\'{o} et al.\ (1998, and references therein). 
We proceeded twofold:  \\

\noindent ~i) The natural magnitude of the filter N is defined as:  \mbox{$\lambda_{Nn} = 
-2.5\cdot log \: (ADU's)_N$}, where $\lambda_N$  stands for the corresponding filters  
$U, B, V, R,$ and $I$, $ADU's$ for the analog-to-digital counts, 
and the subscript $n$ for the corresponding quantity in the natural photometric 
system.  The atmospheric extinction coefficients for a given filter have been estimated by
transforming the nightly $\lambda_{Nn}$'s to the corresponding magnitudes in the standard
system, $\lambda_{Ns}$'s, with the following equation:
\begin{equation} 
\lambda_{Ns} - \lambda_{Nn} = (\rm{zero~point})_{N} - \kappa^{'}_{N}\cdot X_{Nn} -
\kappa^{''}_{N,12} \cdot X_{Nn} \cdot (\lambda_{1} - \lambda_{2})_s 
\end{equation} 
   
\noindent where $X_{Nn}$ is the air-mass when measuring $\lambda_{Nn}$. The 
subscript {\small $N,12$} indicates that the color $(\lambda_1 - \lambda_2)_s$ was used to determine 
the second-order extinction coefficient of filter N.  Here we follow the convention that the effective wavelength  
$\lambda_{1eff} < \lambda_{2eff}$ to construct the color $(\lambda_{1} - \lambda_{2})$.  Finally, 
for a proper determination of \mbox{$\kappa^{'}_{N}$} and \mbox{$\kappa^{''}_{N,12}$} by a least 
square solution, sufficiently large ranges in the airmasses and colors of the standard stars 
($\Delta X_N \ge 0.7 $  and $\Delta (\lambda_1 - \lambda_2)_n \ge 0.8$ for SPM) must be obtained. 
Note that the standard magnitudes and colors are known to an  accuracy of about two percent, 
reflected in the errors of the final transformations of the (bright) standard stars, and that the 
observed magnitudes and airmasses are measured quantities that can have an even better precision. 
To further simplify the equations, the extra-atmospheric instrumental magnitudes 
were then introduced using the extinction coefficients of the night: 
\begin{equation}
\lambda_{Ni} = \lambda_{Nn} - \kappa^{'}_{N} \cdot X_{Nn} - \kappa^{''}_{N,12} 
\cdot X_{Nn} \cdot (\lambda_{1} - \lambda_{2})_s ~.
\end{equation} 

\noindent An instrumental color is the subtraction of two instrumental magnitudes with different passbands, \\  
\mbox{$ (\lambda_{1} - \lambda_{2})_i = \lambda_{1i} - \lambda_{2i}$}. \\

\noindent ii) Once the atmospheric extinction coefficients $\kappa^{'}_{N}$ and $\kappa^{''}_{N,12}$ have 
been determined and applied, the nightly transformation coefficients are calculated 
(i.e. $\beta_{12}$ and $\gamma_{12}$) with the following relations for the colors:
\begin{equation}
(\lambda_{1} - \lambda_{2})_i = \kappa_{0,12} + \beta_{12} \cdot (\lambda_{1}  - \lambda_{2})_s  + 
\gamma_{12} \cdot  (\lambda_{1} - \lambda_{2})^2_s 
\end{equation}  

\noindent Due to the Balmer discontinuity that lies in both the U and B passbands, a better
transformation for the $U$--$B$ color has been achieved by substituting the quadratic term on the right
side of the above equation with a linear term in the color $B$--$V$, obtaining the following expression:  
\begin{equation} 
 (\lambda_1 - \lambda_2)_i = \kappa_{0,12} + \beta_{12} \cdot (\lambda_1 - \lambda_2)_s + 
  \gamma_{12} \cdot  (\lambda_2 - \lambda_3)_s,  
\end{equation}  

\noindent where $\lambda_{1eff}\ < \lambda_{2eff}\ < \lambda_{3eff}$.  For the case of the
magnitude $V$, Equation (3) has been used as follows: 
\begin{equation} 
V_i - V_s   = \kappa_{01} + \beta_{12}\cdot (\lambda_1 - \lambda_2)_s  + \gamma_{12}\cdot 
(\lambda_1 - \lambda_2)_s^2.
\end{equation}  

\noindent For Equations (3)--(5), $\kappa_{0,12}$ and $\kappa_{01}$ are the zero-points of the
transformations of the colors $(\lambda_1 - \lambda_2)_s$, i.e. $U$--$B$, $B$--$V$,
$V$--$R$, $V$--$I$, ..., and of the $V$ magnitude, respectively.  The coefficients
$\beta_{12}$ and $\gamma_{12}$ are the respective first- and second-order transformation
coefficients.  

In general, the second-order atmospheric extinction coefficient $\kappa^{''}_{VR}$ is expected 
to be close to zero due to the nearly constant level (ozone-band contribution) of the atmospheric 
extinction curve at SPM near 5500\AA\ (Schuster \& Parrao 2001).  The second-order extinction 
and linear-transformation coefficients for correcting to extra-atmospheric standard magnitudes 
and colors are very similar from night to night, and also from run to run, because, i) the SPM has
excellent sky conditions, and ii) the same instrumental setup, observing techniques, and data
reduction procedures were used for all nights during both observing runs.  In Table~2 the mean
zero-point corrections, atmospheric extinction, and transformation coefficients are given.

\begin{table*}    
\centering
\caption{Atmospheric extinction and transformation coefficients} \
  {\scriptsize 

\begin{tabular}{cccccccc}
\hline \hline 
color     &$\lambda_1\; \lambda_2\; \lambda_3$& $\kappa^{'}_{1}$ &$\kappa^{''}_{1,12}$& $\kappa_{0,12}$&
$\beta_{12}$&$\gamma_{12}$& rms \\ \hline
          &       &        &        &        &       &        &        \\ 
\multicolumn{8}{c}{July 2001} \\
$(U$--$B)$& $U\,B\,V$  & $0.472\;\,$  & $-0.056\;\,$ & $+$1.625 & 0.711 & $+$0.263   & 0.028  \\ 
$(B$--$V)$& $B\,V\,-$  & $0.243\;\,$  & $-0.050\;\,$ & $+$0.409 & 1.016 & $-$0.050   & 0.010  \\ 
   $V$    & $V\,R\,-$  & $0.106\;\,$  & $+0.079\;\,$ & $+$2.375 & 0.033 & $-$0.008   & 0.016  \\ 
$(V$--$R)$& $V\,R\,-$  & $0.104*$     & $+0.030*$    & $+$0.027 & 0.973 & $+$0.011   & 0.012  \\ 
$(V$--$I)$& $V\:I\:-$  & $0.087*$     & $-0.035*$    & $-$0.151 & 0.923 & $+$0.070   & 0.010  \\ 
          &       &        &         &        &       &        &        \\ 
\multicolumn{8}{c}{February 2002} \\
$(U$--$B)$& $U\,B\,V$  & $0.325\;\,$  & $-0.056\;\,$ & $+$1.765 & 0.751 & $+$0.313   & 0.037  \\ 
$(B$--$V)$& $B\,V\,-$  & $0.212\;\,$  & $-0.050\;\,$ & $+$0.470 & 0.979 & $-$0.023   & 0.018  \\ 
   $V$    & $V\,R\,-$  & $0.082\;\,$  & $+0.079\;\,$ & $+$2.455 & 0.035 & $-$0.054   & 0.027  \\ 
$(V$--$R)$& $V\,R\,-$  & $0.054*$     & $+0.030*$    & $-$0.000 & 1.023 & $-$0.008   & 0.012  \\ 
$(V$--$I)$& $V\:I\:-$  & $0.056*$     & $-0.035*$    & $-$0.165 & 1.038 & $+$0.004   & 0.014  \\ 
\hline 
\end{tabular}
   }
\vspace*{1mm} 

\noindent 
{* indicates that extinction coefficients refer to $\lambda{_2}$, otherwise to $\lambda{_1}$}

\vspace*{3mm}
\end{table*}   

{\bf In Tables~3, 4, and 5 are given the final transformed CCD {\sl UBVRI} photometric values for the open
clusters, Be~89, Ru~135, and Be~10, respectively.  In the text only the first three lines of each table
are shown as examples.  The full versions will be available online at the CDS and WEBDA.  In these tables
the columns 1 and 2 give the following:  X and Y (pixels), the position of a star in the CCD field;
columns 3, 5, 7, 9, and 11: the magnitude and color indices, $V$, $(B$--$V)$, $(U$--$B)$, $(V$--$R)$, and
$(V$--$I)$, respectively (in magnitudes); and columns 4, 6, 8, 10, and 12: the respective photometric errors,
$\sigma_{V}, \sigma_{B-V}, \sigma_{U-B}, \sigma_{V-R}$, and $\sigma_{V-I}$ (in magnitudes), as
provided by IRAF.}

\begin{table*}[!t] 
\caption{}
\vspace*{-2.5mm} 

   {\scriptsize 
\begin{center}
\begin{tabular}{cccccccccccc}
\multicolumn{12}{c}{\large CCD $U\!BV\!RI$  photometry of Be~89} \\[2mm] 
\hline 
X &Y & V &$\sigma_{V}$&(B-V) &$\sigma_{B-V}$ &(U-B) &$\sigma_{U-B}$&(V-R)&$\sigma_{V-R}$&(V-I) &$\sigma_{V-I}$\\
\hline
767.7 &538.3 &11.261 & 0.006 & 0.441 & 0.009 &0.002 &0.007 &0.315 &0.009 &9.999  &9.999 \\
466.5 &496.2 &12.118 & 0.011 & 1.362 & 0.018 &1.191 &0.011 &0.742 &0.013 &1.399  &0.013 \\
511.7 &742.6 &12.172 & 0.008 & 0.443 & 0.011 &0.268 &0.005 &0.296 &0.011 &0.569  &0.017 \\
$\cdot \cdot \cdot$&$\cdot \cdot \cdot$&$\cdot \cdot \cdot$&$\cdot \cdot \cdot$&$\cdot \cdot 
\cdot$&$\cdot \cdot \cdot$&$\cdot \cdot \cdot$&$\cdot \cdot \cdot$&$\cdot \cdot \cdot$&$\cdot 
\cdot \cdot$&$\cdot \cdot \cdot$&$\cdot \cdot \cdot$ \\
\hline
\end{tabular}  

\vspace*{1mm}
\hspace*{-18ex}{\bf Note to table:} The full version of this table is available online at the CDS
and WEBDA. 
\end{center} 
   } 
\end{table*}  


\begin{table*}[!t] 
\caption{}
\vspace*{-2.5mm} 

   {\scriptsize  
\begin{center}
\begin{tabular}{cccccccccccc}
\multicolumn{12}{c}{\large CCD $U\!BV\!RI$  photometry of Ru~135.} \\[2mm] 
\hline 
X &Y & V &$\sigma_{V}$&(B-V) &$\sigma_{B-V}$ &(U-B) &$\sigma_{U-B}$&(V-R)&$\sigma_{V-R}$&(V-I) &$\sigma_{V-I}$\\
\hline
862.7 &231.5 &11.132 &0.003 &0.597 &0.004 &0.212 &0.002 &0.360 &0.010 &9.999  &9.999 \\
237.1 &325.4 &11.689 &0.003 &0.710 &0.003 &0.364 &0.003 &0.387 &0.014 &0.860  &0.011 \\
294.0 &387.3 &11.871 &0.008 &0.742 &0.003 &0.405 &0.004 &0.450 &0.038 &0.895  &0.014 \\
$\cdot \cdot \cdot$&$\cdot \cdot \cdot$&$\cdot \cdot \cdot$&$\cdot \cdot \cdot$&$\cdot \cdot \cdot$&$\cdot \cdot \cdot$&$\cdot \cdot \cdot$&$\cdot \cdot \cdot$&$\cdot \cdot \cdot$&$\cdot \cdot \cdot$&$\cdot \cdot \cdot$&$\cdot \cdot \cdot$ \\
\hline
\end{tabular}  

\vspace*{1mm} 
\hspace*{-18ex}{\bf Note to table:} The full version of this table is available online at the CDS and WEBDA.
\end{center}
\vspace*{1mm}
   } 
\end{table*} 

\begin{table*}[!t] 
\caption{}
\vspace*{-2.5mm} 

   {\scriptsize  
\begin{center}
\begin{tabular}{cccccccccccc}
\multicolumn{12}{c}{\large CCD {\sl UBVRI}  photometry of Be~10.} \\[2mm] 
\hline 
X &Y & V &$\sigma_{V}$&(B-V) &$\sigma_{B-V}$ &(U-B) &$\sigma_{U-B}$&(V-R)&$\sigma_{V-R}$&(V-I) &$\sigma_{V-I}$\\
\hline
374.5 &153.4 &11.244 &0.004 &0.433 &0.004 &0.246 &0.002 &0.220 &0.005 &0.500 &0.004 \\
528.6 &217.9 &12.352 &0.002 &1.483 &0.004 &1.346 &0.009 &0.851 &0.003 &1.631 &0.003 \\
804.6 &881.2 &12.746 &0.003 &0.738 &0.014 &0.231 &0.004 &0.440 &0.014 &0.873 &0.005 \\
$\cdot \cdot \cdot$&$\cdot \cdot \cdot$&$\cdot \cdot \cdot$&$\cdot \cdot \cdot$&$\cdot \cdot \cdot$&$\cdot \cdot \cdot$&$\cdot \cdot \cdot$&$\cdot \cdot \cdot$&$\cdot \cdot \cdot$&$\cdot \cdot \cdot$&$\cdot \cdot \cdot$&$\cdot \cdot \cdot$ \\
\hline
\end{tabular}  

\vspace*{1mm} 
{\bf \hspace*{-18ex}Note to table:} The full version of this table is available online at the CDS and WEBDA.
\end{center} 
   } 
\end{table*}  

\subsection{The data inspection tools ELIPSE and SAFE} 

Since the stellar density of a cluster increases towards its center with respect to the field
stars, an AWK macro \citep[{ELIPSE,}][]{moitinh03} was used to extract the data
of the central region of a given cluster, as defined by visual inspection in a visual (V) or
red (R) image, thus increasing the contrast of the cluster with respect to the surrounding
field stars.  An ellipse was fitted visually to the image in order to extract the photometric
data of the central region of the cluster.  To further support the analyses of the clusters, a
java-based computer program \citep[{SAFE,}][]{mcfarla10} was developed and used to
help us in the visualization and analysis of the photometric data \citep[e.g.][]{schetal07}.
These programs facilitate the elimination of field and apparent non-member stars of a given
cluster from the diagnostic diagrams used to enhance the apperception of cluster features.
Once a satisfactory first estimate of the parameters was obtained, a full-frame solution was
also consulted and refined.

{\sl SAFE} is capable of displaying simultaneously in different color-color (CC) and
color-magnitude (CM) diagrams the cluster's data and has an interactive way to identify a
(group of) star(s) in one particular diagram and to see where it falls in the other diagrams.
This program is capable of displaying up to 16 different diagrams for a given cluster and is
very useful for the determination of a cluster's physical parameters.  Figure~1(a--c)
presents the DSS red-filter images of Be~89 (panel~a), Ru~135 (b) and Be~10 (c), with the
regions analyzed in this work enclosed by ellipses. The central (X,~Y) pixel coordinates of the 
nearly circular regions in Figure~1(a--c), which are considered for the photometric analyses 
are the following: (584, 488) pixels for Be~89, (542, 504) for Ru~135, 
and (517, 493) for Be~10.  The diameters in arcminutes ($\Delta X$, $\Delta Y$) of 
nearly circular regions in Figure~1(a--c) are the following: (2.27, ~2.65) for Be~89, 
(2.62, 2.67) for Ru~135, and (3.12, 2.34) for Be~10.

\section{Analyses of the open clusters Be~89, Ru~135, and Be~10} 

The ({\sl U--B, B--V}), two-color or CC, diagram, and five CM diagrams
have been used together with the zero-age-main-sequence (ZAMS) intrinsic-color
calibrations of Schmidt-Kaler (1982, hereafter SK82) and with the Padova isochrones
(Girardi et al.~2000, hereafter GBBC; Bertelli et al.~2008; Marigo et al.~2008, MGBG)
to obtain reddenings, metallicities, distance moduli, and ages for these clusters.

Our analysis technique for our program clusters places particular emphasis upon the fit of the
ZAMS intrinsic colors and Padova isochrones to the observational data of the clusters, and this
depends in turn upon important characteristics of the CM and CC diagrams for open clusters
\citep[e.g.][their $\S3$]{paunzen06}, which are summarized as follows:
\begin{enumerate} 
 \item
A procedure for eliminating non-members, 
\item
A determination of the interstellar reddening as accurately as possible,
\item
Visibility of the turn-off,
\item
Compensation for binary stars which tend to widen the main-sequence distribution,
\item
Consideration of the red-clump stars (if present) to improve the isochrone fit, and
\item 
An appropriate choice of the isochrone which corresponds to the correct heavy-element
abundance (Z).
\end{enumerate}

Regarding the locus of the main sequence in a CM diagram, and independent of any cosmic
dispersion, the main-sequence strips or bands in the CM diagrams are affected by the
contamination of binary and multiple stars; particularly, the mid-points are shifted to brighter
magnitudes and the colors to redder values due to this contamination, and also sometimes due to
variable intercluster extinction.  For this reason, the SK82 ZAMS and the MGBG isochrones have
been fitted to the blue- and faint-most concentrations within the observed broad main-sequence
bands whenever possible, assuming that these concentrations reflect the single-star distribution
\citep[e.g.][]{bcarney01}, and that most stars observed red- and bright-ward of these are
in fact binary, or multiple, systems.

In the absence of proper-motion/radial-velocity measurements to insure cluster membership,
and to minimize the effects of field-star contamination, we have concentrated more on the
central regions of the clusters rather than using the full-frame CCD images; this has been
accomplished by using the ELIPSE or SAFE programs, described above.  These have been used
to select an elliptical, or polygonal area (with as many as 10 sides), centered on the open
cluster as seen in a $V$ or $R$ image, excluding stars outside this area from further
analyses.  (See Figure~1).  These interactive analyses increase greatly the contrast of
cluster members with respect to the field stars, and thus the scatter in the CM and CC
diagrams is significantly reduced.  

Also, the observational errors, e.g.\ $\,\sigma_{(U-B)}$, of
these three clusters have been considered as a criterion in selecting the more reliable
data for further analyses.  The values of $\sigma_{(U-B)}$ are almost always larger than the
ones of $\sigma_{(B-V)}$ due to the smaller sensitivity of the CCD in the ultraviolet, and
the errors $\sigma_{(R-I)}$, $\sigma_{(V-I)}$, and $\sigma_{(V-R)}$ are among the smallest.
The observational errors, such as $\sigma_{(U-B)}$, $\sigma_{(B-V)}$, and $\sigma_{(B-R)}$,
have been selected to be less than $\approx 0\fm10$ (and sometimes $\la 0\fm05$) in some of
the diagnostic diagrams presented in the analyses to follow, such as the $(U$--$B, B$--$V)$,
$(V,B$--$V)$, and $(V,B$--$R)$ diagrams.

Interstellar reddenings of the program clusters have been estimated from shifts of the intrinsic-color
sequences of SK82 in the {\sl (U--B,B--V)} diagram, until the best fit to the data of the clusters:
along the {\sl U--B} axis by 0.72E{\sl (B--V)}+0.05{\sl E(B--V)}$^2$ and along the {\sl (B--V)}
axis by {\sl E(B--V)}.  For this, F-type stars have been fitted above the main sequence of SK82
(i.e.\ blue-ward in $(U$--$B)$), and simultaneously the red-clump stars above the red-giant colors of
SK82 with consistent ultraviolet excesses according to the normalizations of Sandage (1969).  The
two-color sequence of SK82 has been constructed from the intrinsic colors of SK82 for zero-age
dwarfs ($(B$--$V)_0 \la 0\fm75$) and for giants ($(B$--$V)_0 \ga 0\fm75$). 

Once the two-color sequence of SK82 has been fixed as indicated above, to determine the photometric
metal abundance, [Fe/H], one first locates the F-type stars in the ({\sl U--B,~B--V}) diagram and
compares their location with that of their counterparts of known metallicity (e.g.\ the SK82's ZAMS
calibration).  Deviations between the two are due mainly to their differences in metal content, an
ultraviolet excess, $\delta(U$--$B)$, being caused by differences in line blanketing.  The
metal-deficient F-type cluster stars, if present, lie blue-ward of the ``hump region'' of the ZAMS
sequence, where an eyeball-fitted osculating curve similar to ``the hump'' has been fitted to the
data points of the F-stars (i.e.\ the thick line above the hump of the F-star region in Figures 2, 5,
and 8) and, simultaneously, to the red-clump stars (if present), since they also will lie blue-ward
of the red-giant colors of SK82 with a corresponding ultraviolet excess.  This ultraviolet excess
is correlated with the photometric metallicity of the cluster.  Then, a metallicity value, [Fe/H],
for a cluster can be derived from the empirical calibration, [Fe/H]--$\delta(U$--$B)_{0.6}$, of
Karata\c{s} \& Schuster (2006), allowing the determination of [Fe/H] independently of the
isochrones to be fitted to the data, thus reducing from three to two the free parameters to be
derived from the CM diagrams.  Heavy-element abundances ($Z$) of the three clusters have been
obtained from the photometric metal abundances [Fe/H] with the expression \\ 
\begin{eqnarray}
 Z = Z_\odot \cdot 10^{[Fe/H]},\; Z_\odot = +0.019.
\end{eqnarray} 

\noindent
Finally, the appropriate isochrones of MGBG were computed online in terms of the resulting heavy-element
abundance for further analyses of the clusters (distances and ages).

To estimate the the age of a cluster ($A$) and the true distance modulus \mbox{(DM$= V_0$--$M_{V})$}
in a \mbox{CM diagram}, for example  the ({\sl V, B--V}) diagram, the absolute magnitudes, $M_{V}$,
of the MGBG isochrones have been shifted by $DM +3.1E\: (B-V)$ along the magnitude axis and their
corresponding colors, $(B$--$V)_0$, reddened by adding the color excess $E(B$--$V)$ until some DM value
provides a good fit of the appropriate isochrone to the faint/blue concentration of the observed lower
main sequence of the cluster and, if present, of the red-clump stars.  One has to take into account when
determining the DM that metal-poor stars are sub-luminous as compared with their solar-like counterparts
by determining a reliable value for Z from the CC diagram.  To infer the age of the cluster, the
(logarithm of the) age of the isochrones, $log \: A$, has been varied until a good match with the observed
sequences, i.e.\ the upper main-sequence (MS), the turn-off (TO) stars, and, if present, the red-clump
(RC) stars, has been achieved.  A fine tuning of the DM has been made if necessary. The uncertainties of
$E(B$--$V)$, $Z$, $DM$, and $log \: A$ are discussed in Section~4.

Following a similar procedure to that outlined above, the distance moduli and cluster ages have also been
derived from analyses of four other CM diagrams for each of the clusters.  The corresponding color
excesses applied in the diagrams were iterated starting with the previously derived color excess estimates,
{\sl E(B--V)}, and the results were inter-compared by means of the standard interstellar extinction law
adopted \citep[see Table~6 below; also cf.\ ][]{deaetal78,jmathis90,straizy95} until satisfactory solutions
were obtained for all the CM diagrams.  The derived extinction laws do not differ significantly from that
of Table~6.

\begin{table}[!h] 
\caption{A\lowercase{dopted interstellar extinction law}} 
\begin{center}
\begin{tabular}{cccc}
 \midrule 
$\frac{E(V-I)}{E(B-V)}$&$\frac{E(R-I)}{E(B-V)}$&$\frac{E(V-R)}{E(B-V)}$&$\frac{E(B-R)}{E(B-V)} $\\
\midrule
        1.25           &         0.69          &          0.56         &           1.56         \\ 
\midrule
\end{tabular} 
\end{center}  
\end{table}  

\subsection{Be~89} 

The ({\sl U--B, B--V}) diagram of Be~89 is shown in Figure~2.  An interstellar reddening  of
$(B$--$V) = 0\fm60 \pm 0\fm09$ has been derived by shifting the intrinsic two-color stellar sequence of SK82
along the reddening vector as described in the previous section. (Another possibility, to fit the
stars by $E(B$--$V) \simeq 0\fm73$ to the blue (B-star) branch of the ZAMS curve, would leave many stars
far from a good fit).  Six stars apparently in the cluster are noticed with $(B$--$V) \approx 1\fm6$ and
$(U$--$B) \approx 1\fm4$  (big open circles in Figure~2) lying near, but above (i.e.\ blue-ward of) the
giant sequence of SK82, the expected location of the RC stars; their subsequent locations in the CM diagrams
confirm this classification (see Figures~3 and 4).  A seventh candidate falls further from the expected RC
locus in four of the five CM diagrams.

The F-type and RC stars of Be~89 (cf.\ Figure~2, $(B$--$V) \approx 1\fm0$ and $\approx 1\fm6$, respectively)
lie above the (reddened) ZAMS two-color calibration of SK82 by $\delta(U$--$B)\simeq 0\fm1$.  Our best
eyeball fit to the data is shown as the heavy solid curve in Figure~2.  In the dereddened two-color
diagram, the heavy line gives a value of $(U$--$B)_{0}= -0\fm10\pm 0\fm02$ at $(B$--$V)_{0} = 0\fm44$, and at
this same color index, the highest point of the SK82 hump has $(U$--$B)_{0} = -0\fm02$.  The resulting
ultraviolet excess, $\delta(U$--$B)=+0\fm08\pm 0\fm02$, has been converted to $\delta(0.6) = +0\fm10\pm
0\fm02$ at $(B$--$V) = +0\fm60$ with the normalization ratios given by Sandage (1969, his Table 1A).  These
values have been listed in Table~7, together with the corresponding photometric metallicity [Fe/H]$
= -0.35\pm 0.02$ dex derived with help of the calibration \mbox{[Fe/H]--$\delta(0.6)$} of Karata\c{s}
\& Schuster (2006). Note that the $\delta(0.6)$ in the notation of the latter authors corresponds to
delta(0.6) in the notation of Sandage (1969).  Applying the above relation between [Fe/H] and $Z$,
where [Fe/H] has been estimated as $-0.35\pm 0.02$ dex, gives $Z=+0.008\pm0.0003$.  The online
isochrones of MGBG have been iterated using this metal abundance when further analyzing Be~89.

\begin{table*}[!h] 
\caption{Normalized (U--B) excesses and derived metallicities}
\vspace*{-2.5mm}
\begin{center}
\begin{tabular}{lcccccc}
\toprule 
Cluster& $(U$--$B)_{SK82}$ &$(U$--$B)_{0,fit}$& $\delta(U$--$B)$ &$\delta(0.6)$& [Fe/H]   &  Z   \\
       &     [mag]         &     [mag]        &      [mag]       &    [mag]    & [dex]    &      \\
\midrule 
Be~89  &     $-0.02$       &    $-0.10$       &      0.08        &     0.10    &$-0.35$   & 0.008 \\ 
Ru~135 &     $-0.02$       &    $-0.16$       &      0.14        &     0.14    &$-0.71$   & 0.004 \\
Be~10  &     $-0.02$       &    $-0.09$       &      0.07        &     0.11    &$-0.49$   & 0.006 \\[-4mm] 
       &                   &                  &                  &             &                  \\ 
error  &$\le \pm 0.01$     &$\pm 0.02$        &$\pm 0.02$        &$\pm 0.02$   &$\pm 0.02$&$\le \pm 0.001$ \\
\bottomrule 
\end{tabular} 
\end{center} 
\end{table*}   

\begin{figure}[!t] 
\includegraphics[width=1.25\columnwidth]{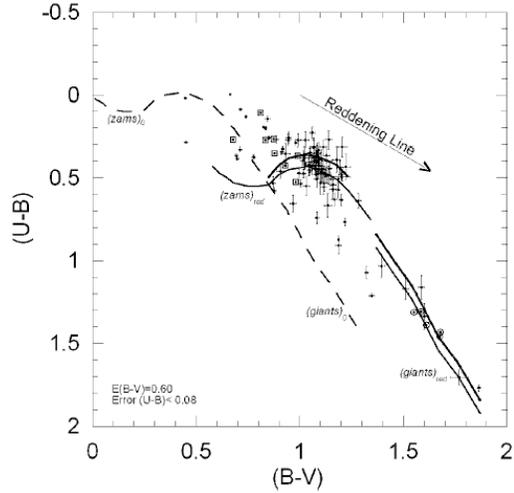}
\caption {The $(U$--$B, B$--$V)$ diagram of Be~89.  The ``S'' curves (upper
parts, ZAMS, and lower parts, red giants) have been taken from the two--color
relations of SK82 and are displayed for the interstellar reddening values
$E(B$--$V)=0\fm00$ and  $0\fm60$ (the bluer and redder versions,
respectively).  A reddening vector is also shown as an arrow, and big open
circles mark the six RC candidates. {\bf and open squares, the blue-straggler
ones}. A heavy solid curve represents our
best fit to the data which has been adjusted to the main-sequence F-type
stars above (i.e.\ blue-ward of) the ZAMS colors of SK82 and, simultaneously,
to the RC stars above the red-giant colors of SK82.  This fit has been used
to estimate the heavy-element abundance of the cluster, which is shown in
Table~7.}
\end{figure} 

\begin{figure}[!b] 
\includegraphics[width=0.95\columnwidth]{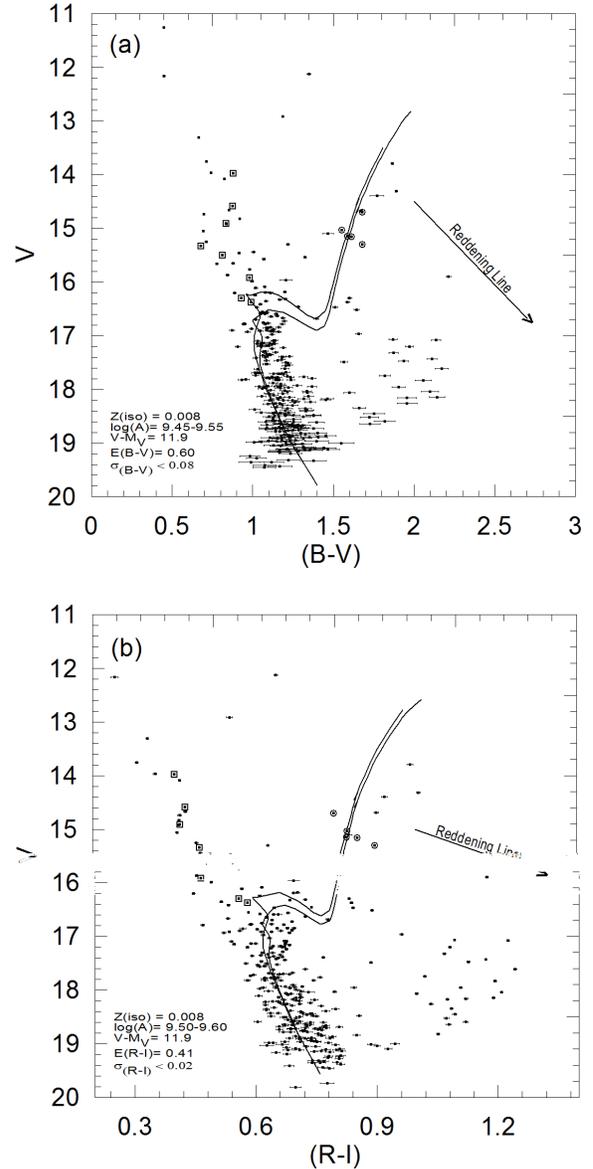} 
\caption {CM diagrams, $(V,B$--$V)$ and $(V,R$--$I)$, for Be~89.
Solid lines show the interpolated $Z=+0.008$ isochrones of MGBG (cf.\ Table~7 for
the inferred metallicity).  Big open circles denote the RC candidates, and open
squares, the blue-straggler ones.  See the text and Table~8 for the inferred
values of the distance modulus and age.
    }
\end{figure} 

\begin{figure}[!t] 
\includegraphics[width=0.8\columnwidth]{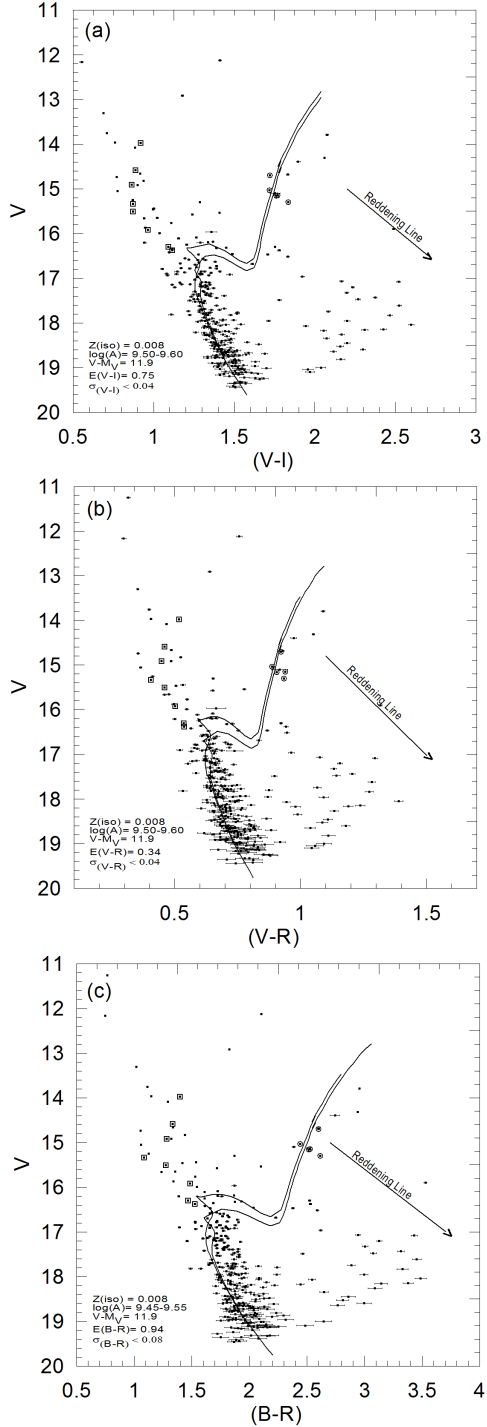}
\caption {CM diagrams, $(V,V$--$I)$, $(V,V$--$R)$, and $(V,B$--$R)$,
(top, middle and bottom panels, respectively) for Be~89.  The isochrone curves
and the symbols have the same meaning as in Figure~3.  See the text and Table~7
for the inferred values of reddening and metallicity, and Table~8 for the distance
modulus and age.}
\end{figure} 

\begin{figure}[!t] 
\includegraphics[width=1.02\columnwidth]{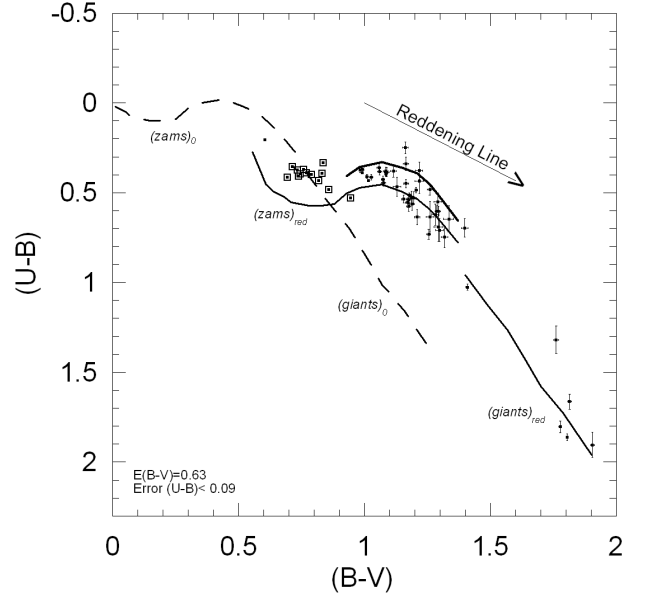}
\caption {The CC diagram of Ru135.  The SK82 curves and the
symbols have the same meaning as in Figure~2.  See the text and Table~7
for the inferred values of the reddening and metallicity.
      }
\end{figure} 

In Figures~3 and 4, the isochrones of MGBG for $Z=+0.008$ have been over-plotted 
in five CM diagrams:  $V,(B$--$V)$, $V,(R$--$I)$, $V,(V$--$I)$, $V,(V$--$R)$, and
$V,(B$--$R)$ after reddening the isochrones along the color axis with
a color excess corresponding to $E(B$--$V)=0\fm60$, converted with help of Table~6,
and adding a visual extinction of $A_{\rm V} = 3.1\cdot E(B$--$V) = 1\fm 86$ to
the absolute magnitudes of the isochrones.  The isochrones have then been shifted
vertically to obtain the best fit to the observed lower-MS and and RC sequences.
This vertical shift is the (true) distance modulus, $DM =(V_{0}$--$M_{\rm V})$.
The best fit for Be~89 is  $DM = 11\fm90\pm 0\fm06$ (d$ = 2.4\pm 0.06$ kpc, cf.\
Table~8).

To derive an age estimate for Be~89, the isochrones of MGBG for $Z = +0.008$ have
been shifted in the CM planes as above, i.e.\ $ M_{V}+3.1E(B$--$V)+DM $ and
$C_0(\lambda_1-\lambda_2) + E[C(\lambda_1-\lambda_2)]$, respectively, where the
latter color excesses have been computed from $E(B$--$V)$ with help of Table~6, and
then the isochrone ages have been varied until a satisfactory fit to the data has
been obtained through the observed upper-MS, TO, and RC sequences of the cluster
(cf.\ Figures~3--4).  The resulting inferred mean age of Be~89 is
$\log(A) = 9.58\pm 0.06\;$ dex $( A = 3.8 \pm 0.6 $ Gyr).

For all of these CM diagrams of Be~89, two isochrones have been plotted to provide
a means for appreciating the uncertainties of the derived distances and ages.  In
Table~8 the range in ages provided by these isochrone pairs, the final values for
the distances and ages from each CM diagram, and the mean values for each cluster are
given.  Error estimates of $(V_{0}$--$M_{\rm V})$ and $log (A)$ are discussed in $\S3.4$
below, and the mean results given in Table~8 have been calculated with Equations~(8)
and (9) inserting the corresponding parameters summarized in the table.

\begin{figure}[!p] 
\includegraphics[width=0.95\columnwidth]{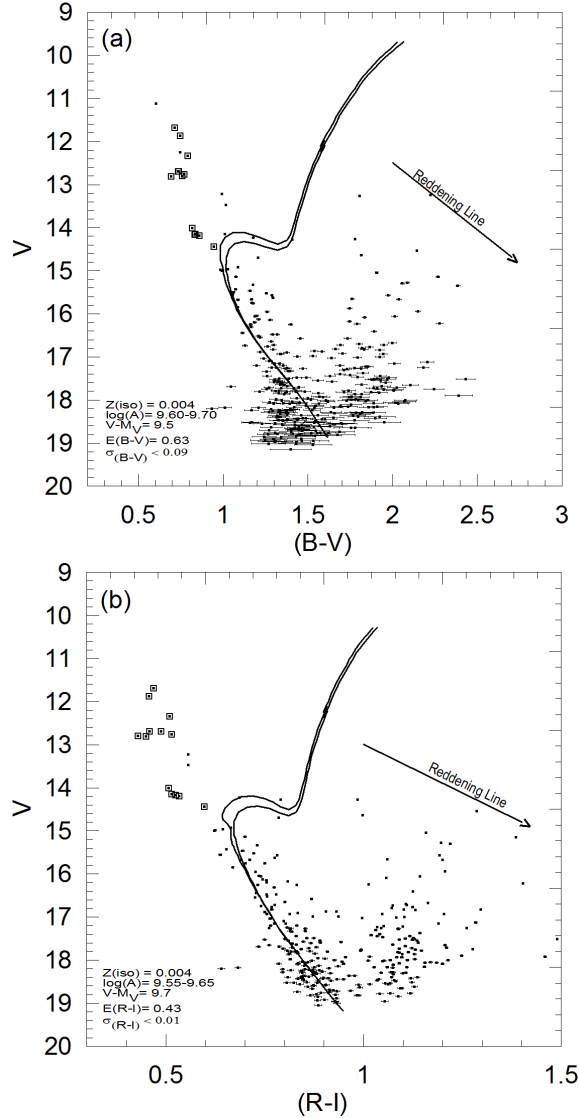}
\caption {The $(V,B$--$V)$ and $(V,R$--$I)$ diagrams for Ru~135.  Solid lines
show the isochrones of MGBG interpolated to $Z=+0.004$.  See the text, and Tables~7
and 8, for the inferred values of reddening, metallicity, distance modulus, and age.
Stars shown with open-square symbols are most likely field, or blue-straggler, stars.}
\end{figure} 

\subsection{Ru~135}  

The same procedures outlined in $\S3$, and $\S3.1$\ for Be~89, have also been used
for the clusters Ru~135 and Be~10.  A reddening of $E(B$--$V) = 0\fm63 \pm
0\fm12$ has been derived for Ru~135 (cf.\ Figure~5).  However, a clump of A-type stars
at $(B$--$V) \simeq 0\fm8$ and $(U$--$B)\simeq 0\fm4$ seems to be present, with a
horizontal-like distribution which does not fit satisfactorily the reddened two-color
ZAMS curve of SK82.  These stars (Sp $\approx$ A-types) are probably less reddened
than Ru~135 by $\simeq 0\fm3$ in $E(B$--$V)$, nearer, and most probably not cluster
members (cf.\ the open squares in the CC and CM diagrams of Figures~5, 6, and 7), or
they could be blue stragglers belonging to the cluster.  For this latter case, they
would be peculiar because of an ultraviolet-flux excess present in their spectral
energy distributions (SEDs), and only a spectroscopic study with good signal-to-noise
ratios would reveal more about their true nature. 

\begin{figure}[!t] 
\includegraphics[width=0.75\columnwidth]{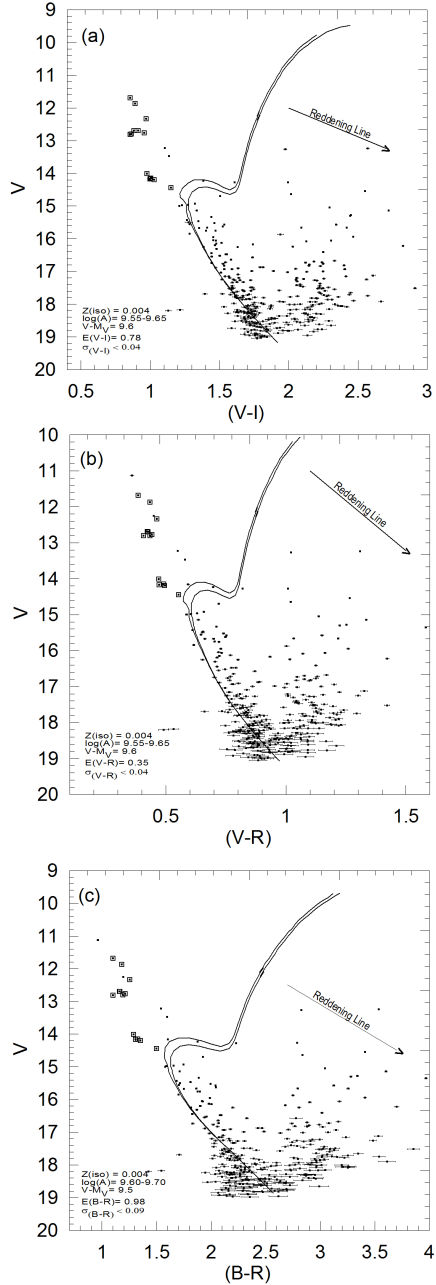}
\caption {The $(V,V$--$I)$ , $(V,V$--$R)$ and $(V,B$--$R)$ diagrams
for Ru~135.  The symbols are the same as in Figure~6.  See the text, and Tables~7 and 8,
for the inferred values of reddening, metallicity, distance modulus, and age.}
\end{figure} 

Ru~135 contains a considerable number of F- and later-type stars, and appears to
have its blue-most turn-off limit at $(B$--$V) \approx +1\fm0$ and $(U$--$B) \approx 0\fm4$,
corresponding to a dereddened $(B$--$V) \approx +0\fm43$ (i.e.\ Sp $\approx$ F5V).  The best
fit to the observed F-hump sequence in the $(U$--$B, B$--$V)$ diagram is the solid curve
shifted blue-ward with respect to the two-color SK82 curve (cf.\ Figure~5).  From the
ultraviolet excess of these cluster F stars and following the procedure outlined at the 
beginning of $\S3$, [Fe/H]$ = -0.71\pm 0.02$ dex ($Z = +0.004\pm 0.0002$) has been derived.
The isochrones of MGBG with this metallicity have been computed on line and used in the
following analyses.

\begin{table*} \centering   
  \vspace*{-2.5mm}
  \setlength{\tabnotewidth}{0.5\columnwidth}
  \tablecols{6}
  \caption{Distance and age estimates of the clusters} 
\begin{tabular}{lccccc} 
\toprule 
Color        &$(V_{0}$--$M_{V})$&      d        &   $log(A)$   &   $log(A)$   &    $A$        \\
             &     [mag]      &    [kpc]      &   range    &              &   [Gyr]       \\
\midrule 
             &                &               &            &              &               \\[-3mm]
 &\multicolumn{4}{c}{\underline{Be~89: ~E(B-V)$ = 0.60\pm 0.09$ ~\&~ $Z = +0.008\pm 0.001$}} & \\ 
{\sl (B--V)} & $11.90\pm 0.10$  & $2.4\pm 0.1$  & 9.45--9.55 & $9.55\pm 0.15$  & $3.6\pm 1.4$  \\
{\sl (R--I)} & $11.90\pm 0.20$  & $2.4\pm 0.2$  & 9.50--9.60 & $9.60\pm 0.20$  & $4.0\pm 2.3$  \\
{\sl (V--I)} & $11.90\pm 0.15$  & $2.4\pm 0.2$  & 9.50--9.60 & $9.60\pm 0.20$  & $4.0\pm 2.3$  \\
{\sl (V--R)} & $11.90\pm 0.10$  & $2.4\pm 0.1$  & 9.50--9.60 & $9.60\pm 0.15$  & $4.0\pm 1.6$  \\
{\sl (B--R)} & $11.90\pm 0.20$  & $2.4\pm 0.2$  & 9.45--9.55 & $9.55\pm 0.10$  & $3.6\pm 0.9$  \\
             &                &               &            &              &               \\[-2mm]
  Mean       &$11.90\pm 0.06$ &$2.4\pm 0.06$ &            &$9.58\pm 0.06$& $3.8\pm 0.6$  \\ \hline
             &                &               &            &              &               \\[-2mm]
&\multicolumn{4}{c}{\underline{Ru~135: ~E(B-V)$=0.63\pm 0.12$ ~\&~ $Z=+0.004\pm 0.001$}}  \\
{\sl (B--V)} & $~9.50\pm 0.15$  & $0.75\pm 0.05$& 9.60--9.70 & $9.60\pm 0.15$ & $4.0\pm 1.6$  \\
{\sl (R--I)} & $~9.70\pm 0.15$  & $0.87\pm 0.06$& 9.55--9.65 & $9.55\pm 0.15$ & $3.6\pm 1.5$  \\
{\sl (V--I)} & $~9.60\pm 0.20$  & $0.83\pm 0.08$& 9.55--9.65 & $9.55\pm 0.15$ & $3.6\pm 1.5$  \\
{\sl (V--R)} & $~9.60\pm 0.15$  & $0.83\pm 0.06$& 9.55--9.65 & $9.60\pm 0.15$ & $4.0\pm 1.5$  \\
{\sl (B--R)} & $~9.50\pm 0.20$  & $0.75\pm 0.07$& 9.60--9.70 & $9.60\pm 0.15$ & $4.0\pm 1.6$  \\
             &                &               &            &              &               \\[-2mm]
  Mean       &$9.58\pm 0.07$  & $0.81\pm 0.03$&            &$9.58\pm 0.06$&$3.8\pm 0.7$   \\ \hline
             &                &               &            &              &               \\[-2mm]
 &\multicolumn{4}{c}{\underline{Be~10: ~E(B-V)$=0.75\pm 0.09$ ~\&~ $Z=+0.006\pm 0.001$}}  \\
{\sl (B--V)} & $11.20\pm 0.11$  & $1.7\pm 0.1$  & 9.05--9.15 & $9.05\pm 0.10$& $1.1\pm 0.3$  \\
{\sl (R--I)} & $11.10\pm 0.20$  & $1.7\pm 0.2$  & 9.10--9.20 & $9.10\pm 0.10$&  $1.3\pm 0.3$  \\
{\sl (V--I)} & $11.15\pm 0.15$  & $1.7\pm 0.1$  & 9.05--9.15 & $9.05\pm 0.15$&  $1.1\pm 0.3$  \\
{\sl (V--R)} & $11.15\pm 0.20$  & $1.7\pm 0.2$  & 9.05--9.15 & $9.05\pm 0.10$&  $1.1\pm 0.3$  \\
{\sl (B--R)} & $11.20\pm 0.10$  & $1.7\pm 0.1$  & 9.05--9.15 & $9.05\pm 0.05$& $1.1\pm 0.1$  \\
             &                &               &            &              &               \\[-2mm]
  Mean       &$11.16\pm 0.06$ &$1.70\pm 0.05$ &            &$9.06\pm 0.05$& $1.08\pm 0.08$\\
\bottomrule 
\end{tabular} 
\end{table*} 

The five CM diagrams, $(V,B$--$V)$ through  $(V,B$--$R)$, of Ru~135 are displayed in
Figures~6 and 7 together with the reddened isochrones that best fit the data for the
derived color excess and metallicity, $E(B$--$V)= 0\fm63$ and $Z = +0.004$.
The distance moduli, $(V_{0}$--$M_V)$, and ages, $A$, found from these five CM diagrams
and their respective isochrone fittings are given in Table~8.

In these CM diagrams a significant number of stars are seen extending to brighter magnitudes
and red-ward from the fainter and redder observational limits of the main sequences,
i.e.\ the stars extending red-ward and upward from $(V,B$--$V) \approx (18\fm 5,1\fm5)$ or
$V,(R$--$I) \approx (18 \fm 5, 0\fm9)$ (cf.\ Figure 6).  These are probably field
red-giant stars contributed by the Galactic bulge, as suggested by the Galactic longitude
and latitude of Ru~135, $\ell \simeq 16.4^{\circ}$ and $b \simeq +6.2^{\circ}$
\citep[see][Figures~3.5 and 2, respectively]{jbinney98,staetal96}.  The fact that Ru~135
lies near the direction of the Galactic central region also explains the significant number
of brighter and bluer foreground stars seen in its CC and CM diagrams.

\begin{figure}[!t] 
\includegraphics[width=1.02\columnwidth]{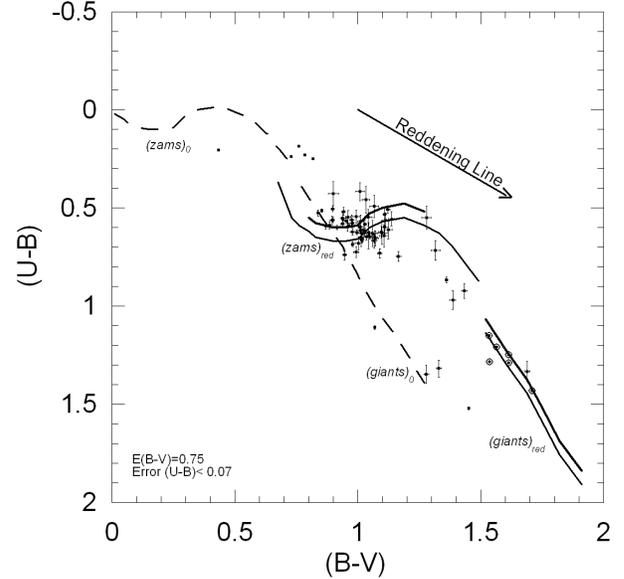} 
\caption {The $(U$--$B, B$--$V)$ plot of Be~10.  The symbols and the curves are the same
as in Figure~2.}
\end{figure} 

\begin{figure}[!p]  
\includegraphics[width=0.8\columnwidth]{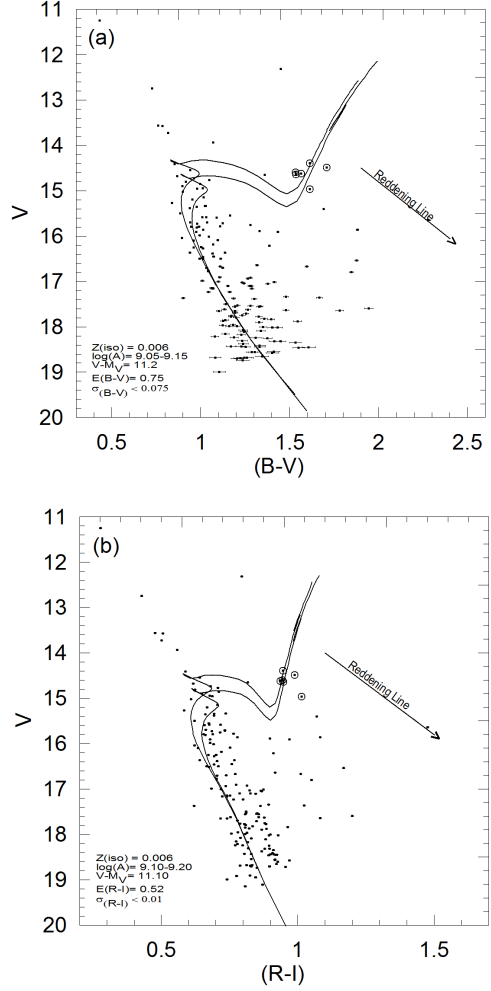} 
\caption {The $(V,B$--$V)$ and $(V,R$--$I)$ diagrams for Be~10.
Solid lines show the isochrones of MGBG interpolated to $Z=+0.006$.  The larger open circles
identify the RC candidates.  See the text, and Tables 7 and 8, for the inferred values for
reddening, metallicity, distance modulus, and age.}
\end{figure} 

\begin{figure}[!ht]  
\includegraphics[width=1.3\columnwidth]{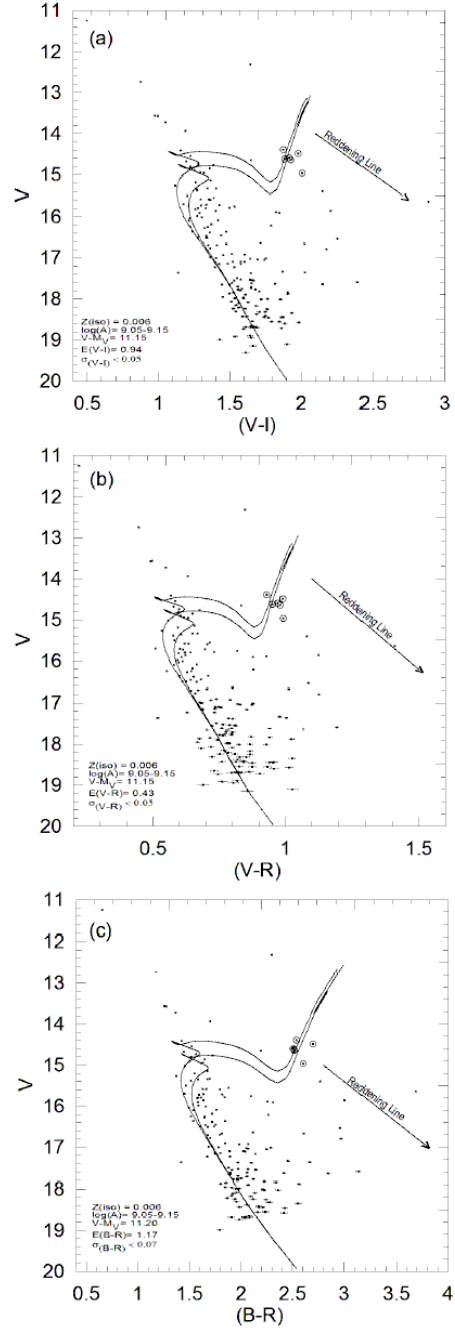}
\caption {The $(V,V$--$I)$, $(V,V$--$R)$ and $(V,B$--$R)$
diagrams for Be~10.  The symbols are the same as in Figure~9.
See the text, and Tables 7 and 8, for the inferred values for reddening,
metallicity, distance modulus, and age.}
\end{figure} 

\subsection{Be~10}  

In Figure~8 the loci of stars observed in the direction of Be~10 are shown in the $(U$--$B,B$--$V)$
diagram, together with the standard interstellar reddening vector and the two-color curve of SK82,
shifted along this vector to procure the best fit to the data.  From the fits along the $(B$--$V)$
and $(U$--$B)$ axes, $E(B$--$V) = 0\fm75 \pm 0\fm09$ and ([Fe/H], $Z) = (-0.49\pm 0.02$
dex, $+0.006\pm 0.0003)$ are found, following the procedures described in 
$\S3$ and $\S3.1$ (see Table~8 for partial and mean results).  Again, the appropriate isochrones of
MGBG have been computed online with this corresponding metallicity and are used below for further
analyses of Be~10.

For Be~10, $DM = (V - 3.1 \cdot E(B$--$V) - M_{V}) = 11\fm16\pm 0\fm06$, the distance, $d = 1.7\pm 0.05$
kpc, the metallicity, $Z = +0.006\pm 0.0003$, $log(A) = 9.06\pm0.05$, and the age, $A= 1.08 \pm 0.08$ Gyr
have been measured.  Our results are listed in Tables~7 and 8.  The resulting (best) isochrone fitting to
the corresponding Be~10 data in the $(V,B$--$V)$, $(V,R$--$I)$, $(V,V$--$I)$, $(V,V$--$R)$ and
$V,(B$--$R)$ diagrams are displayed in Figures~9 and 10, where one can  see that the isochrones
reproduce well the observed lower and upper MS, the TO, and RC sequences of this cluster.

\vspace{0.2cm}

\subsection{Estimated errors and weighted averages}  

\vspace{0.1cm}

In Table~7, the ultraviolet excesses and the metallicities are given for Be~89, Ru~135, and
Be~10, and in Table~8, the mean values for the distance moduli, heliocentric distances,
logarithmic ages, and ages, together with the corresponding estimates of precision.  The errors
were calculated in a straightforward manner \citep[cf.][and references therein]{bevingt03}.
In the following the details of this error analysis are presented.

\subsubsection{Errors in $E(B$--$V)$ and $Z$}  

The random errors in the color excess $E(B$--$V)$ and photometric metallicity [Fe/H] were
estimated as follows: 
\begin{itemize} 

\item[~~i)] By moving the two-color curve of SK82 backward and forward along the standard
reddening vector in the $(U$--$B,B$--$V)$ diagram until a good fit with the observed
GK-type,  RC, and F-hump sequences was achieved. (No BA-type sequences were present for
these clusters.)  The precision of the determinations also depends on the scatter of
the data points (cf.\ Figure~5 of Ru~135 and Figure~8 of Be~10 for good and regular cases,
respectively).  The uncertainties given in Tables~7 and 8 reflect this.  Following this
procedure, a typical error in $E(B$--$V)$ for the quality of the reduced data of our
clusters is (conservatively) $\approx 0\fm04$.  The systematic error in $E(B$--$V)$ depends
on the color calibration used.  In the case of SK82, the uncertainty can be safely
assumed to be, at the most, of the order of the difference between two adjacent spectral
subclasses.

\item[~ii)] The random photometric-metallicity uncertainty has then been estimated from
the parabolic (eyeball) fit to the data of the maximum characterizing the ultraviolet flux
excess of the stars at the dereddened color \mbox{$(B$--$V)_0$} $ \simeq 0\fm44$
(Sp $\simeq$ F5) and then following Sandage's (1969) normalization procedure.

The uncertainty of the metal content $Z$ was determined from the relation
\citep[e.g.][]{bevingt03}: 

\begin{equation} 
 \sigma_Z = ln\:10\times Z\times \sigma_{[Fe/H]}. 
\end{equation}

\noindent 
$\sigma_{[Fe/H]}$ has been estimated from the uncertainty in the ultraviolet excess $\delta
(U$--$B)$ at the F hump between the observed and the SK82 two-color curves and is
typically $\pm 0\fm02$.  Assuming $\langle Z \rangle = 0.006$ (the mean of the three clusters)
$\mid\!\!{\sigma_Z}\!\!\mid \leq 0.0003$ is obtained with Equation (7) above.  Assuming
an error of about 0.001 for $Z$ is, in our case, a quite conservative estimate.

\item[iii)] On the other hand, the deviation of the assumed reddening vector from the ``true''
one depends on the quotient $\frac{E(U-B)}{E(B-V)}$, which can {\bf strongly}
deviate locally from its canonical value of 0.72 \citep[see][and references therein]{chaetal87,
johnson77}.  This uncertainty may produce errors larger than the precision quoted above.
For a crude estimate, using the extremes of the cited values of $\frac{E(U-B)}{E(B-V)}$ and a
typical color excess of $E(B$--$V) = 0\fm50$, the uncertainty in $\delta (U$--$B)$ could be as
large as 0\fm15--0\fm20.  However, since our displacements of the SK82 curve in the CM-diagrams are
consistent with the canonical value for the interstellar extinction law, we assume that this
error contribution is negligible statistically.

\item[~iv)] Another systematic uncertainty results from the two-color calibration of SK82:
from the uncertainty of $(U$--$B)$ or $(B$--$V)$ in the two-color calibration curve of SK82,
which is expected to be of the order of the difference between two typical spectral subclasses
(in our case $\lesssim 0\fm05$) for the (U--B) index; and from the fit of the whole curve to a
cluster data set, of the order $\lesssim \frac{0\fm05}{\sqrt{N}}$, where there are N  pivotal
points considered when adjusting the SK2 two-color curve (the BA-type, the F-hump,
the GK-type, and the RC sequences; i.e.\ $N\gtrsim 3$).  Summarizing, the systematic uncertainty
in $\delta (U$--$B)$ should be less than roughly three times that given by the precision of the
flux measurements.
\end{itemize} 

\vspace{-0.1em}

\subsubsection{Errors in the distance moduli and ages}  
\begin{itemize}

\item[~~i)] The uncertainties $\sigma_{DMi}$ for the moduli given in Table~8 that result from fitting the
appropriate isochrones to the data in the CM diagrams depend on the photometric
uncertainties (flux measurement and standard-transformation errors), the absolute
magnitude and intrinsic color calibration errors (see for example, SK82), the color excess
uncertainty of a given color (which depends on $E(B$--$V)$), and on the reddening law adopted.  We
assume that the isochrones only contain the errors of the absolute magnitude and intrinsic-color
calibrations and that the photometric and transformation errors are small ($\approx 0\fm03$) when
compared to the other sources of error.  In our case, the largest contribution to the distance
modulus uncertainty is due to the uncertainty in the absolute-magnitude scale, followed by the
uncertainty in the slope of the reddening vector, and  the color-excess error, about 0\fm3, 0\fm15
and 0\fm12, respectively, which combine to give an expected total uncertainty as large as $\sigma_{DMi}
\cong 0\fm25$.

The moduli resulting from the CM diagrams of each object and the mean moduli for the three clusters
are given in Table 8, and the mean of the moduli has been derived from the five moduli, weighted
with their respective (usually unequal) precisions, with the following expression:
\begin{equation}
\overline{DM} = \frac{\Sigma (DM)_i/\sigma^2_{DMi}}{\Sigma [1/\sigma^2_{DMi}]};\; (i=1,...,5), 
\end{equation}
and the associated mean uncertainty is estimated from the individual uncertainties of the five
CM-diagrams of a given cluster by the relation:
\begin{equation}
 \frac{1}{(\sigma_{mean})^2} = \sum {\frac{1}{(\sigma_{DM})^2_i}} 
\end{equation}

\noindent
The combined error is the square root of the sum of the squared uncertainties and is expected to be
about $0\fm15$, or even less.

\item[~ii)]The uncertainty in the $log(A)$ has a random error due to the (eyeball) fit of the
isochrone with the appropriate metallicity to a given CM diagram of a cluster in question, and a
quantitative estimate is obtained by jiggling bright-ward and faint-ward the isochrone curve until
a good fit of the lower main sequence produces the $DM$.  Then the age of the isochrone
is varied until a good fit to the upper main sequence, the TO, and the RC sequences is achieved.
The two isochrones shown in the CM diagrams of the program clusters give a quantitative estimate of
this last error.  Several different authors have computed isochrones as function of the metallicity,
and the physics behind seems to be well understood.  One does not expect a large variation in the 
$log(A)$ error due to any uncertainty in the physics, and the uncertainties of $E(B$--$V)$ and
$(V_0$--$M_V)$ play a secondary role because the age errors depend more on the form of the
isochrone curve and how it embraces the data (i.e.\ the the upper main-sequence and TO regions
and the RC sequence) and, less significantly, on the reddening law (except perhaps the blue and
near-ultraviolet filters).  More problematic is the case when the TO region is not well defined
(i.e.\ isolated from field stars) and/or the RC sequence is not present.  In our case, the
errors for the different colors of Table~8 reflect these uncertainties.
\end{itemize}

\section{Comparison of fundamental parameters of the three clusters} 

The reddening values of our three clusters have been compared to ones derived
from the dust maps of Schlegel, Finkbeiner \& Davis (1998; hereafter, SFD);
these are based on the COBE/DIRBE and IRAS/ISSA maps, and take into account
the dust absorption all the way to infinity.  $E(B$--$V)(\ell, b)_\infty$ values
of our three clusters have been taken from SFD maps using the web pages of
NED\footnote[6] {http://nedwww.ipac.caltech.edu/forms/calculator.html}.  These
$E(B$--$V)(\ell, b)_\infty$ values are $0\fm99$ for Be~89, and $1\fm06$ for
both Ru~135 and Be~10.  However, Arce \& Goodman (1999) caution that SFD
reddening maps overestimate the reddening values when the color excess
$E(B$--$V)$ is more than $\approx 0\fm15$.  For the revision of SFD reddening
estimates,  the equations of Bonifacio, Monai \& Beers (2000) and Schuster et
al.~(2004) have been adopted. Then the final reddening, $E(B$--$V)_{A}$, for a
given star is reduced compared to the total reddening $E(B$--$V)(\ell, b)_\infty$
by a factor $\lbrace1-\exp[-d \sin |b|/H]\rbrace$, where $b$, $d$, and $H$ are
the Galactic latitude (Column~2 of Table~9), the distance from the observer to
the object (Column~7 of Table~9), and the scale height of the dust layer in the
Galaxy, respectively; here we have assumed $H = 125$ pc (Bonifacio, Monai \& Beers
2000).  Note that Galactic latitudes of our three clusters are less than
$10^\circ$.  These reduced final reddenings are $E(B$--$V)_{A} = 0\fm54$ for Be~89,
$0\fm36$ for Ru~135, and $0\fm64$ for Be~10.

For Be~89, our reddening value of $0\fm60$ is in good agreement with the value
of $0\fm54$ obtained from the dust maps of SFD.  For Be~10 our reddening value of
$E(B$--$V)=0\fm75$ is within about $1\sigma$ of the value $0\fm64$ derived from
the SFD dust maps, and for Ru~135, our reddening value of $0\fm63$ differs by about
$2\sigma$ from the value of $0\fm36$ obtained from these SFD maps.
These reddening values derived by different methods are in reasonable agreement
with each other, giving confidence to our results.

As can be seen from the summarized results given in Table~9, the reddening value
$0\fm 60$ found here for Be~89 is smaller than the $E(B$--$V)=1\fm03$ of Tadross (2008;
hereafter T08a), and than the $E(B$--$V)=1\fm05$ of \citep[][hereafter S10]{subrama10}.
Our derived distance modulus and distance for Be~89, ($(V_{0}$--$M_{V}), d$(kpc)) =
$(11\fm90\pm 0\fm06, 2.4\pm 0.06)$, are smaller than the values of $(12\fm39, 3.00)$ of
T08a and larger than the $(11\fm54, 2.04)$ of S10.  Our inferred age
$(\log(A), A(Gyr))$ = $(9.58, 3.8$ Gyr) for this cluster is considerably older than
$(8.93, 0.85$ Gyr) given by T08a and larger than the estimate $(9.02, 1.06\;$ Gyr)
by S10.  For the analysis of Be~89, T08a used $JHK$ photometry and the
isochrones of Bonatto et al.~(2004) with  a solar metallicity.  This is, partially,
the origin of the disagreement between the two age estimates, since our lower metallicity
for Be~89 will necessarily lead to a larger age for a given TO.  Also, most probably,
the differences are partially due to the different procedural approaches for
estimating the fundamental parameters; we derive in a straightforward manner the
estimates for the interstellar extinction and metallicity:   by fitting SK82's ZAMS
to the data in the $(U$--$B,B$--$V)$ diagram, by then measuring the ultraviolet excess
of the F-type stars to derive a cluster metallicity, and finally using the appropriate
isochrones in CM diagrams to estimate the true distance modulus and age of Be~89.  Two
parameters (reddening and metallicity) are estimated in a CC diagram separately from
the other two parameters (distance and age) from the CM diagrams.  S10 have also
assumed a solar metallicity ($Z_\odot$) for their isochrones (from GBBC) and have
used only CM diagrams to estimate the reddening, distance, and age of Be~89.

Previous results in the literature for Ru~135 are found in the work by
Tadross (2008; hereafter T08b), and for Be~10 in the papers by Lata et al.
(2004; L04) and Maciejewski \& Niedzielski (2007; MN07).  Our
reddening value $E(B$--$V)= 0\fm63\pm 0\fm 12$ for Ru~135 is significantly
smaller compared to the reddening value of $1\fm10$ given by T08b.  Also,
our derived distance modulus and distance, ($(V_{0}$--$M_{V}), d$(kpc)) =
$(9\fm58, 0.81)$, for Ru~135 are significantly smaller than
$(11\fm33, 1.85)$, and our inferred age $(\log(A), A)$ =
$(9.58, 3.80$ Gyr) is considerably older than $(8.70, 0.50$ Gyr),
values by T08b.

In defense of the present results, our value for $E(B$--$V)$ ($0\fm63$)
falls between the value derived from SFD ($0\fm36$) and the value of T08b
($1\fm10$); our value is in much better agreement with SFD.  T08b used the
solar-metallicity isochrones of Bonatto, Bica, \& Girardi (2004), and his
results are based on the comparison of isochrones to observed data in the
$J,(J$--$H)$ and $K,(J$--$K)$ planes of infrared photometry.  These
differences between our values and those of T08b are probably due mainly to
the largely different values for the interstellar reddening, but also to
the difference in the assumed metallicities, to the use of different
stellar models and isochrone sets, which make use of differing input
physics and colour-temperature transformations, and to distinct photometric
data sets.

For the Be~10 open cluster, our reddening value $E(B$--$V)= 0\fm75$ is in
reasonable agreement with the value of $E(B$--$V)= 0\fm87$ given by L04,
and in good agreement with $E(B$--$V)= 0\fm71$ by MN07.  For the
metallicity of the Be~10 cluster, L04 adopt the $Z = +0.008$ isochrones of
Girardi et al.\ (2002), and  MN07 adopt the solar isochrones of Bertelli et
al.\ (1994).  From our two-color diagram,  $Z=+0.006$ has been derived (see
$\S3.3$), which is in agreement, within the error bars, with the value of
L04.  Our distance modulus and distance for Be~10, ($(V_{0}$--$M_{V}),
d$(kpc))$ = (11\fm16, 1.70)$ differ from the values $(11\fm8, 2.3)$ of L04,
but very little from the values of MN07, $(11\fm26, 1.79)$.  Our inferred age
$(\log(A),~A)$ = $(9.06, 1.08$ Gyr) for this cluster disagrees by almost a
factor of two (0.5 Gyr) with L04, but is in good agreement with MN07,
$(9.00, 1.00$ Gyr).  Again, our interstellar reddening for Be~10,
$E(B$--$V)= 0\fm75$ falls between the value derived from SFD ($0\fm64$) and
the value $0\fm87$ by L04.

The age values in Table~9 have been compared to ages estimated with the
(age, $\Delta V$) calibration given by Carraro \& Chiosi (1994; their
equation~(3)).  Note that this last calibration does not consider the metal
abundance of the cluster.  Here, $\Delta V$ means the magnitude difference
between the RC and TO, which is well known as an age indicator.  Both
open clusters Be~89 and Be~10 have RC candidates (see the CM plots for 
these two clusters, Figures~3--4, and Figures~9--10, respectively).
TO values occur at $V \approx 16\fm 5$ for Be~89 and $V \approx 14\fm 8$ for
Be~10, whereas the RCs occur at $V \approx 15\fm3$ and $V \approx 14\fm 7$,
respectively.  From this age--$\Delta V$ calibration of Carraro \& Chiosi
(1994), ages have been estimated as $\log(A)= 9.1$ (1.3 Gyr) for Be~89
and $\log(A)= 8.6$ (0.4 Gyr) for the Be~10.

The average age values given by us, $\log(A)= 9.58$ (3.8 Gyr) for Be~89 and
$\log(A)= 9.06$ (1.08 Gyr) for Be~10 are somewhat older than the ones
estimated from this relation of Carraro \& Chiosi (1994).  However, these
age differences are at least partially explained by the sub-solar
metallicities of these two clusters ([Fe/H]$= -0.35$ dex for Be~89 and
$-0.49$ dex for Be~10; see $\S3.1$, $\S3.3$, and Table~9).  Lower
metallicities require larger ages for the same TO.

\begin{table*}     
\caption{Fundamental parameters of Be~89, Ru~135, and Be~10 }
\begin{minipage}{190mm}
{
\scriptsize

\begin{tabular}{lcccccccccl}
\hline
Cluster& $(l^{\circ},~b^{\circ})$&$E(B$--$V)$&[Fe/H]&$Z$&($V_{0}$--$M_{V})$& $d$ &$\log(A)$&Isochrone& $R_{GC}$ & Reference     \\
       &                         & [mag]     & [dex]&   &    [mag]         &[kpc]&         &         &   [kpc]  &               \\
\hline
\multicolumn{ 1}{l}{{Be~89}}  &~83.16,~+4.82&0.60 &$-0.35$&+0.008 & 11.90  &2.40 &  9.58   &m8$^\dag$&   8.55   & this work     \\
\multicolumn{ 1}{l}{{}}       &~~~~~,~~~~~  &1.03 &$  -  $& solar & 12.40  &3.00 &  8.93   &b4       &   $-$    & Tadross 2008a   \\
\multicolumn{ 1}{l}{{}}       &~~~~~,~~~~~  &1.05 &$  -  $& solar & 11.54  &2.04 &  9.02   &g0       &   $-$    & Subramaniam et al. 2010\\
\hline
\multicolumn{ 1}{l}{{Ru~135}} &~16.42,~+6.23&0.63 &$-0.71$&+0.004 & ~9.58  &0.81 &  9.58   &m8       &   7.72   & this work     \\
\multicolumn{ 1}{l}{{}}       &~~~~~,~~~~~  &1.10 &$  -  $& solar & 11.33  &1.85 &  8.70   &b4       &   $-$    & Tadross 2008b   \\
\hline
\multicolumn{ 1}{l}{{Be~10}}  &138.62,~+8.88&0.75 &$-0.49$&+0.006 & 11.16  &1.70 &  9.06   &m8       &   9.84   & this work     \\
\multicolumn{ 1}{l}{{}}       &~~~~~,~~~~~  &0.87 &$  -  $&+0.008 & 11.80  &2.30 &  8.80   &g2       &   $-$    & Lata et al. 2004 \\
\multicolumn{ 1}{l}{{}}       &~~~~~,~~~~~  &0.71 &$  -  $& solar & 11.26  &1.79 &  9.00   &B4       &   $-$    & MN07       \\
\hline
\end{tabular} 
\vspace*{1mm} 

\noindent 
{\bf $^\dag$Isochrone sources:} B4 = Bertelli et al.~(1994); b4 = Bonatto et al.~(2004); 
g0 = GBBC; g2 = Girardi et al.~(2002); m8 = MGBG
}
\end{minipage}
\end{table*}       

In Table~9 our results are summarized for Be~89, Ru~135, and Be~10: Columns~1 and 2 
contain the cluster name and Galactic coordinates, respectively.  The resulting 
reddening, $E(B$--$V)$, is given in Column~3.  The metallicity and heavy-element 
abundances, [Fe/H] and (Z), are given in Columns~4 and 5, respectively.  True distance 
modulus values, ($V_{0}$--$M_{V})$, and their corresponding heliocentric distances 
to the observer are given in Columns 6 and 7, respectively.  Column 8 gives the
average age (i.e.\ $\log(A)$; where $A$ is in years), as derived from the five
CM diagrams.  Different isochrones used by us and by other authors are
referenced in Column~9.  Average Galactocentric distances are listed in Column~10.
The corresponding references from the literature are listed in Column~11.

\section{Conclusions} 

CCD $U\!BV\!RI$ photometry of three poorly studied Galactic open clusters, Be~89,
Ru~135, and Be~10, has been analyzed, based on new SPM observations.  The fundamental
parameters of reddening, metallicity, age, and distance of these open clusters have
been inferred and presented in Tables~7--9.

The interstellar reddenings and metallicities of these three clusters have been
determined from two-color, $(U$--$B,B$--$V)$, diagrams prior to the use of the CM
diagrams.  Heavy element abundances, $Z$, of the three clusters have been found from
the ultraviolet excess, $\delta (U$--$B)$, of the F-stars by comparison with the
two-color curve of SK82 ($Z_{SK82} = Z_\odot$), by using the normalizations of
Sandage (1969), and by applying the calibration, \mbox{[Fe/H]--$\delta(0.6)$}, of
Karata\c{s} \& Schuster (2006), with the advantage of reducing by two the number of
free parameters of the isochrones when fitting to the data in the CM diagrams.  When
necessary, we have iterated slightly afterwards for a better, more consistent,
solution for the four cluster parameters (reddening, metallicity, distance, and age).
Deeper U frames would improve our determinations employing this method, which allows
us to estimate the reddening and metallicity independently using a CC diagram, in
contrast to the exclusive fitting of isochrones to CM diagrams and the use of the solar
metallicity, which are the more common techniques used in the literature.

The present adjustments of the SK82, CC relations to the MS and RC stars, and
of the MGBG isochrones to MS, TO, and RC stars in the CM diagrams show
good consistency and appropriate fits for all three open clusters, in the one CC
diagram and all five CM diagrams.  Good consistency is seen in the Figures~2--4 for
Be~89, Figures~5--7 for Ru~135, and Figures~8--10 for Be~10.

The CC and CM diagrams of Be~89 and Ru~135 suggest that they are metal-poor and old
for their location in the Galaxy, compared to other open clusters.

For Be~89, stars with $V < 16\fm2$ and $(B$--$V)\leq 0\fm9$ are most likely foreground or
blue-straggler stars.  The blue-straggler and RC candidates in the field of Be~89
need spectroscopic and/or astrometric observations to test their cluster membership and
to elucidate their nature.

Similar candidates for blue-straggler or bright foreground stars are seen in the
CC and CM diagrams of Ru~135 and Be~10, Figures 5--7 and 8--10, respectively.
In the case of Ru~135 and for stars with $V$ fainter than about $14\fm2$, 
the onset of the cluster in the CM diagrams is clearly seen.  Objects brighter
than this limit and with $(B$--$V) \le 0\fm9$ are probably blue stragglers or
bright foreground stars.

Despite its similar age to Be~89, no RC stars are noticeable in the CM diagrams of
Ru~135.  On the other hand, the CC and CM diagrams of Be~10 show clear evidence for
a RC grouping, although it is somewhat younger than the other two clusters.  The lack
of any RC stars in the CM diagrams of Ru~135, contrasting with Be~89 and Be~10, may
result either from relative differences in mass segregation and our emphasis on the
inner regions of these clusters, or from the poorness of these cluster fields and
small-number statistics.  Ru~135, being closer to the Galactic center, may be more
perturbed and less dynamically relaxed than the other two clusters.  Also, Be~89 and
Be~10 each show only eight, or fewer, RC candidate stars, and it is not clear that
all of these are in fact cluster members.

For the typical accuracy of photometric observations (and we are no exception),
the final error estimates are fixed by the accuracy of the cluster parameters as given by
the systematic uncertainties in the absolute-magnitude, intrinsic-color, and
reddening-vector calibrations, for example, the adequacies, or not, of the SK82 colors,
the MGBG isochrones, and the standard interstellar-reddening curve.

Finally, further radial velocity and proper motion information for these clusters 
will allow us to clean with more assurance most non-members from the CC and CM diagrams
in order to obtain better determinations of their physical parameters and to better
understand the nature of the blue-straggler and red-clump candidates in these three
open clusters.  Deeper photometric observations, especially in the $U$ and $B$ bands,
will provide clearer, cleaner, and more precise solutions from the CC diagram.

\acknowledgements 

This work was supported by the CONACyT projects 33940, 45014, 49434 and PAPIIT-UNAM 
IN111500 (M\'exico). {\.I}A acknowledges a grant from the Mexican government
(Secretar\'{\i}a de Relaciones Exteriores).
YK acknowledges financial support of the Scientific and Technical Research Council
of Turkey (TUBITAK, BIDEB-2219).
This research made use of the WEBDA open cluster database of J.-C. Mermilliod.
We also thank an anonymous referee for valuable suggestions and comments that
helped improve this work substantially.

\section*{Supplementary material}
The following material is available online at the CDS and WEBDA: (a) This manuscript.
{\bf (b) Table~3, which presents standard $U\!BV\!RI$ CCD photometry and observing errors for
the open cluster Be~89; the columns 1 and 2 give the following:  X and Y (pixels),
the position of a star in the CCD field; columns 3, 5, 7, 9, and 11:  the magnitude and color
indices, $V$, $(B$--$V)$, $(U$--$B)$, $(V$--$R)$, and $(V$--$I)$, respectively (in magnitudes);
and columns 4, 6, 8, 10, and 12:  the respective photometric errors, $\sigma_{V},
\sigma_{B-V}, \sigma_{U-B}, \sigma_{V-R}$, and $\sigma_{V-I}$ (in magnitudes), as
provided by IRAF.  (c) Tables~4 and 5, which have the same format as Table~3, but for
Ru~135 and Be~10, respectively.}

\end{document}